\documentclass[11pt]{article}\usepackage{graphicx}

\begin{document}

%%------------------------------------------------------

\def\epsfgb#1#2{\includegraphics[width=#2]{#1.eps}}

\def\half{{\textstyle{1\over2}}}
\def\i{{\rm i}}
\def\d{{\rm d}}
\def\e{{\rm e}}
\def\q{{\rm q}}
\def\h{{\scriptscriptstyle{1\over2}}}
\def\pol#1{\vec{#1}}
\def\pmb#1{\setbox0=\hbox{#1}
\kern-.025em\copy0\kern-\wd0
\kern.05em\copy0\kern-\wd0
\kern-.025em\raise.0433em\box0}
\def\mbi#1{{\pmb{\mbox{\scriptsize ${#1}$}}}}
\def\bm#1{{\pmb{\mbox{${#1}$}}}}
\def\strut{\vrule width0pt height 15pt depth 7pt}

\title%
%*
{Excitation of non-quark degrees of freedom in N*\thanks{%
Talk delivered by B. Golli
at the workshop on 
\textit{Few body problems in hadronic and atomic physics} 
        at Bled, Slovenia, July 7--14, 2001.}
}

\author{%
B. Golli$^a$\thanks{%
E-mail: Bojan.Golli@ijs.si},
P. Alberto$^{b}$ and 
M. Fiolhais$^{b}$
\\ 
%}\institute{%
\small
$^a$Faculty of Education, University of Ljubljana, and
	         J. Stefan Institute, Ljubljana, Slovenia 
\vspace*{2mm}\\ \small
$^b$Department of Physics and Centre for Computational Physics, 
          University of Coimbra, 
\\ \small    
P-3004-516 Coimbra, Portugal}

%<
\date{}

%>
%\authorrunning{B. Golli, P. Alberto and M. Fiolhais}
%\titlerunning{Excitation of non-quark degrees of freedom in N*} % optional

\maketitle

\begin{abstract}
We study the contribution of glueball and $\sigma$-meson degrees of
freedom  to nucleon excited states in the framework of a chiral 
version of the chromodielectric model.
We have found that these degrees of freedom considerably lower the
energy of the Roper and may substantially weaken the electroexcitation 
amplitudes for the N(1440) and in particular for the N(1710).
\end{abstract}

\noindent
Among the excited states of the nucleon the Roper resonance, N(1440), 
plays a rather special role since, due to its relatively low 
excitation energy, a simple picture in which one quark populates 
the 2s level does not work here. 
The relatively low energy of the N* has been explained by

\begin{itemize}

\item the residual interaction originating from chiral mesons 
      exchange~\cite{Plessas},

\item allowing the confining potential to vibrate (e.g. in the MIT bag 
  model~\cite{guiMIT}, or in models with dynamically generated confinement,
  \cite{broRPA}),
  
\item  describing the N* as a pure sigma meson excitation rather 
  than the excitation of the quark core~\cite{Krehl}, 

\item  explicit excitations of glue-field~\cite{Li}.

\end{itemize}

We present a simple model, the chromodielectric model (CDM), which 
is particularly suitable to describe the interplay of glueball and 
meson excitations together with quark radial excitations. 
The model includes a chromodielectric field $\chi$ which assures quark 
dynamical confinement, and the chiral fields, $\sigma$ and $\pi$.
The Lagrangian takes the form \cite{BirseP}:
\begin{equation}
  \mathcal{L} = \mathrm{i}\bar{\psi}\gamma^\mu \partial_\mu\psi
  + {g\over{\chi}}\, \bar{\psi}
    ({\hat{\sigma}}+\mathrm{i}\pol{\tau}\cdot\hat{{\pol{\pi}}}\gamma_5)\psi
  + \mathcal{L}_{\sigma,\pi} 
  +  \mathcal{L}_\chi\;,
\label{Lagrangian}
\end{equation}
where
\begin{eqnarray}
  \mathcal{L}_\chi &=&  
  \half\partial_\mu{\hat{\chi}}\,\partial^\mu{\hat{\chi}}
  - {1\over2}M^2\,{\hat{\chi}}^2\;,
\nonumber\\
  \mathcal{L}_{\sigma,\pi} &=&
  \half\partial_\mu{\hat{\sigma}}\partial^\mu{\hat{\sigma}}
  + \half\partial_\mu{\hat{\pol{\pi}}}\cdot\partial^\mu{\hat{\pol{\pi}}} 
  - \mathcal{U}(\hat{{\pol{\pi}}}^2+{\hat{\sigma}}^2)\;,
\end{eqnarray}
and $\mathcal{U}$ is the usual Mexican hat potential.
The model parameters $g=0.03$~GeV and $M=1.4$~GeV are chosen to reproduce 
best the ground state properties~\cite{delta1}; for $m_\sigma$ we 
have taken {$0.6\;{\rm GeV}<m_\sigma<1.2\;{\rm GeV}$}.

The Roper resonance is described as a cluster of three quarks in 
radial-orbital configuration (1s)$^2$ (2s)$^1$, surrounded by pion and 
$\sigma$-meson clouds and by a chromodielectric field.
The fields oscillate together with quarks.
The ans\"atze for the nucleon $|N\rangle$ and the Roper $|R'\rangle$
are assumed in the form of coherent states~\cite{GR85} on top of {$(1s)^3$} 
and {$(1s)^2\;(2s)^1$} configurations, respectively, projected onto 
subspace with good angular momentum and isospin.
Different profiles are assumed for the Roper and the nucleon, and
the boson fields are allowed to adapt to the quark configuration.
The `potential' breathes together with the quarks as illustrated
in Fig.~1.
The proper orthogonalization of both states is ensured 
by writing: 
\begin{equation}
   |{\rm {R}}\rangle 
    = {1\over\sqrt{1-c^2}}(|{\rm {R}}'\rangle - c|{\rm {N}}\rangle)\;,
\qquad
  c = \langle{\rm {N}}|{\rm {R}}'\rangle \;.
\label{orthoR}
\end{equation}
\vspace*{-5mm}
\begin{figure}[ht]
\centerline{\hbox{} \kern-8mm%\kern-10mm
%\epsfgb{figbled1x}{74mm}\kern-6mm\epsfgb{figbled2x}{74mm}\kern-10mm}
\epsfgb{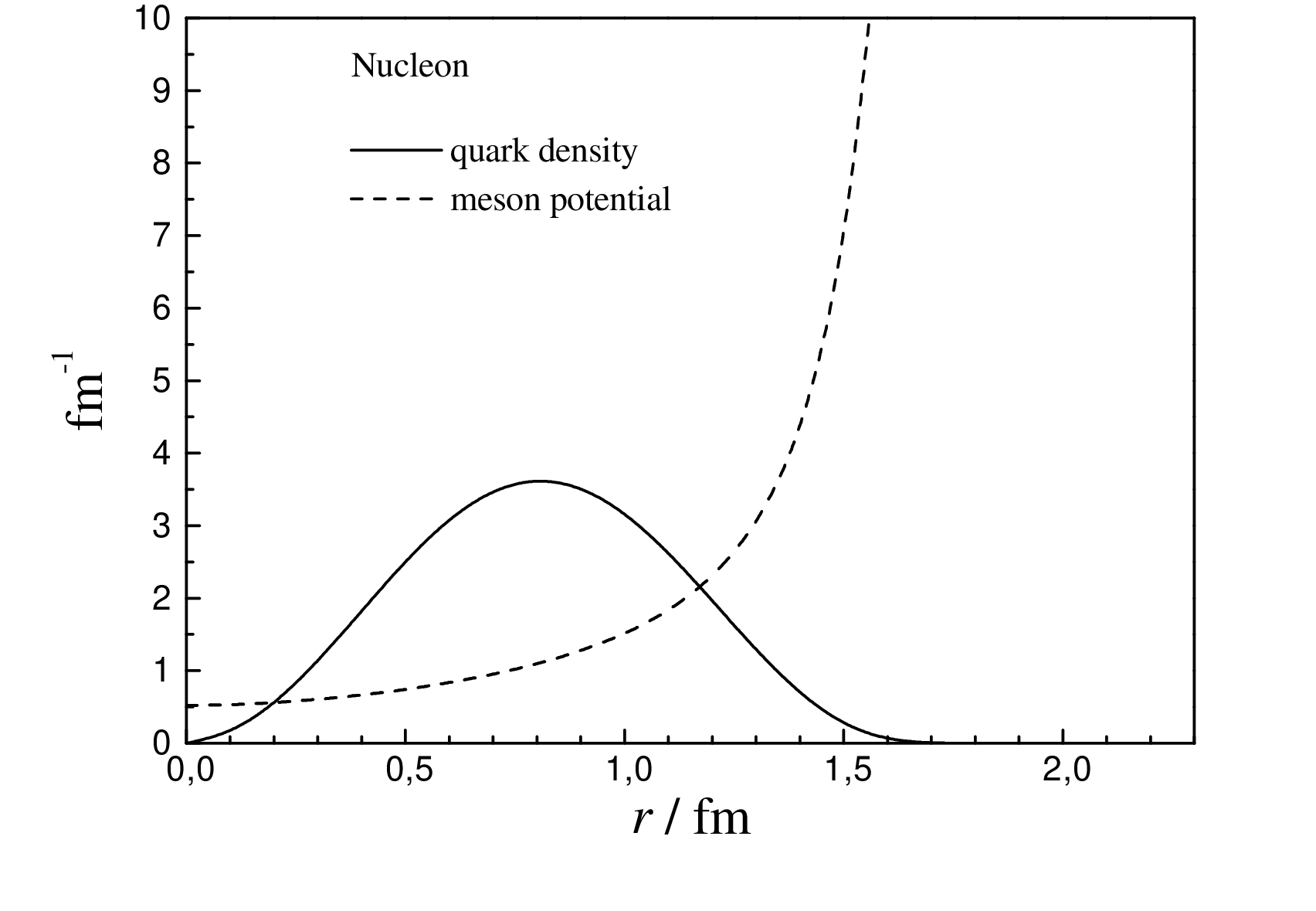}{83mm}\kern-6mm\epsfgb{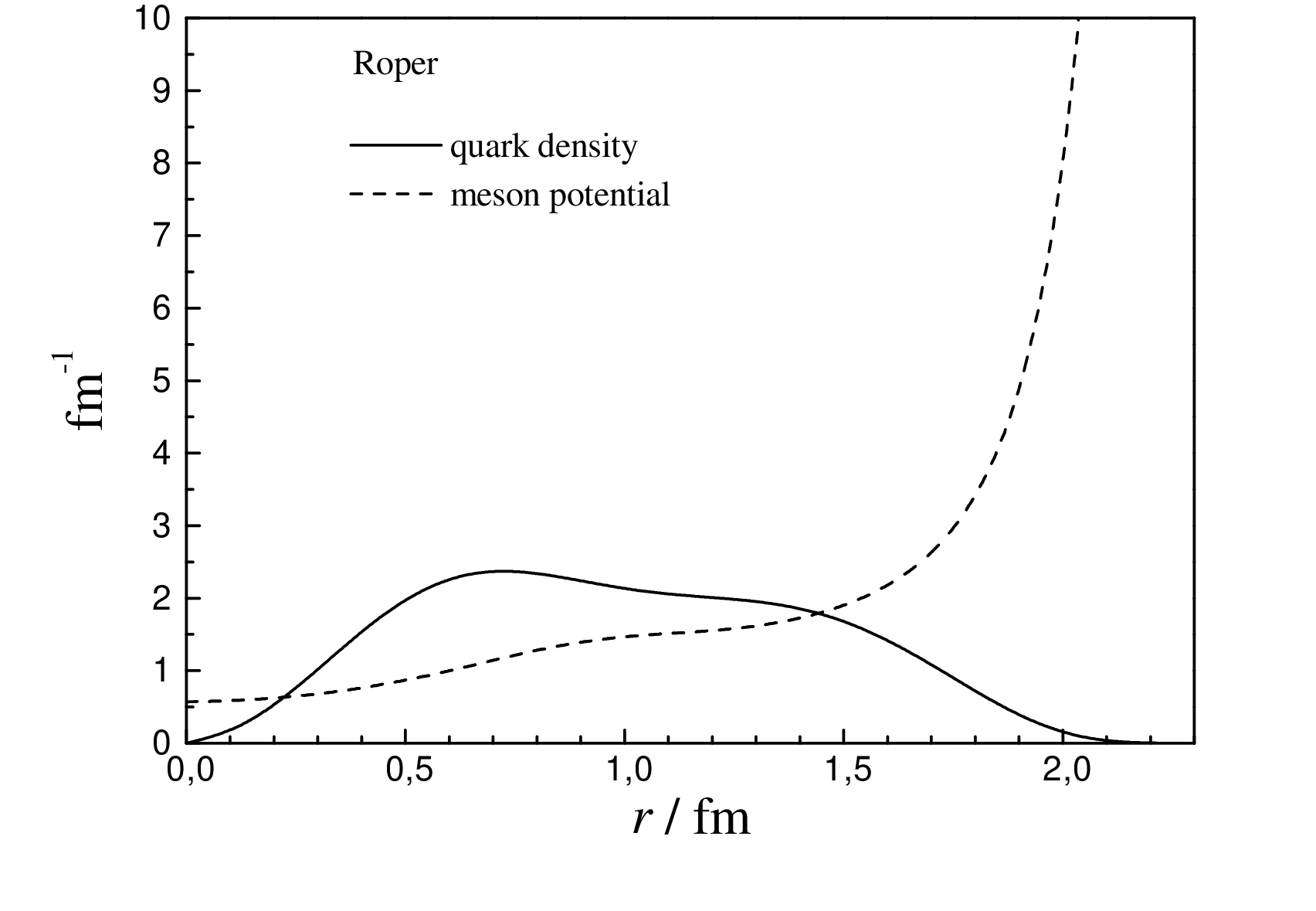}{83mm}\kern-10mm}
\vspace*{-3mm}
\caption{The baryon densities
(solid lines) and the effective potentials (dashed lines) generated 
by the self-consistently determined $\pi$, $\sigma$ and $\chi$ fields
in the nucleon and the Roper.}
\end{figure}

We investigate here another possible type of excitation in which 
the quarks remain in the ground state configuration (1s)$^3$ while 
the chromodielectric field and the $\sigma$-field oscillate. 
Such oscillations can be described by expanding the boson fields as 
small oscillations around their ground state values.
For the $\sigma$-field we write:
$$
  \hat{\sigma}(\vec{r}) =
     \sum_n{1\over\sqrt{2\varepsilon_n}}\, \varphi_n(r)
     {1\over\sqrt{4\pi}}
     \left[{\tilde{a}}_n + {\tilde{a}}^\dagger_n\right] + \sigma(r)\, ,
$$
where
$\varphi_n$ and $\varepsilon_n$
satisfy the Klein-Gordon equation
\begin{equation}
 \left(-\nabla^2 + m^2 
     + {\d^2 {V}({\sigma}(r))\over\d{\sigma}(r)^2}\right)\varphi_n(r)
  = \varepsilon_n^2 \varphi_n(r)\;.
\label{KG}
\end{equation}
A simple ansatz for the anni\-hi\-la\-tion 
(creation) operator of the $n$-th mode is given by
\begin{equation}
 {\tilde{a}}_n = \int\d k\, \tilde{\varphi}_n(k)
    \left({\tilde{a}}(k) 
    - \sqrt{2\pi\omega_\sigma(k)}\,\eta(k)\right)\;,
\qquad
{\tilde{a}}_n|N\rangle = 0\;,
\label{an}
\end{equation}
where $\eta(k)$ and $\tilde{\varphi}_n(k)$ are the Fourier 
transforms of $\sigma(r)$ and $\varphi_n(r)$, respectively, 
and $\omega_\sigma(k)=k^2+m_\sigma^2$. 
The effective potential in (\ref{KG}) is
\begin{equation}
 V_{\sigma\sigma}(r) 
  = {\d^2 V(\sigma(r))\over\d\sigma(r)^2}
  = \lambda \left[C_2\phi(r)^2 
        + 3\sigma(r)(\sigma(r)+2\sigma_v)\right] \, ,
\label{Vsigma}
\end{equation} 
where $\sigma$ is the fluctuating part of the full field and
$C_2$ is a projection coefficient slightly smaller than unity.
Similar expressions hold for the $\chi$ field with
the effective potential
\begin{equation}
 V_{\chi\chi}(r) 
  = {\d^2 V(\chi(r))\over\d\chi(r)^2}
  = -{3\over2\pi}\, {g\over\chi(r)^3}
   \left[(\sigma(r)+\sigma_v)(u(r)^2-v(r)^2) 
        + 2\phi(r)u(r)v(r)\right]\, .
\label{Vchi}
\end{equation} 
The effective potential turns out to be repulsive for the $\chi$-field 
and attractive for the $\sigma$-field; in the latter case there exists 
at least one bound state with the energy $\varepsilon_1$ of typically 
100~MeV below the $\sigma$-meson mass.

The ansatz for the Roper can now be simply extended as
\begin{equation} 
  |{\rm {R}}^*\rangle = c_1|{\rm {R}}\rangle 
       + c_2{\tilde{a}^\dagger}_\sigma|{N}\rangle\;,
\label{Rext}
\end{equation} 
where $\tilde{a}^\dagger_\sigma$ is the creation operator 
for the lowest vibrational mode.
The coefficients $c_i$ and the energy are determined by solving 
the generalized eigenvalue problem in the $2\times2$ subspace.
The solution with the lowest energy corresponds to the Roper,
while the orthogonal combination to one of the higher excited 
states with nucleon quantum numbers, e.g., the $N(1710)$, provided 
the $\sigma$-meson mass is sufficiently small.
The energy of the Roper is reduced (see Table~\ref{Tab:1}),
though the effect is small due to  weak coupling between the 
state (\ref{orthoR}) and the lowest vibrational state with 
the energy $\varepsilon_1$.
The reduction becomes more important if the mass of the
$\sigma$-meson is decreased.
The energy of the combination orthogonal to the ground state 
is close to $E_N+\varepsilon_1$ with
$\sigma$-meson vibrational mode as the dominant component.

\begin{table}[h]
\begin{center}
\begin{tabular}{rrrrrrr}
\hline

\hline
\strut $m_\sigma$ & $\ \ \ \ E_N$ \ & \ \ 2s--1s& 
\ \ \ \ $\Delta E_R$& \ \ \ \ $\Delta E_{R*}$& 
\ \ \ \ \ \  $c_2$& $\ \ \ \ \ \ \varepsilon_1$ \\ 
\hline
\strut 1200 & 1269 & 446 & 354  & 353 & 0.05 & 1090 \\
\strut  700 & 1249 & 477 & 367  & 364 & 0.12 &  590 \\
\hline

\hline 
\end{tabular}
\end{center}
\caption{For two $\sigma$-masses we show the nucleon energy ($E_N$),
the Roper-nucleon energy splittings calculated from the single 
particle energy difference (2s--1s), the state (\ref{orthoR}) 
($\Delta E_R$) and the state (\ref{Rext}) ($\Delta E_{R*}$). 
%The single particle energy difference 2s-1s is obtained using the
%ground states boson fields also for the Roper.
All energies are given in MeV.}
\label{Tab:1}
\end{table}

The electromagnetic nucleon-Roper transition amplitudes
as well as the transition amplitudes to higher excitations with
nucleon quantum numbers represent an important test which may 
eventually distinguish between the models listed at the beginning. 
The transverse helicity amplitude is defined as
\begin{equation}
A_{1/2}=-{\zeta}\,\sqrt{2 \pi \alpha \over k_W}  \int \d^3 {\bm r} \, \
  \langle\tilde{\rm {R}}_{+{1\over 2}, M_T}|{\bm J}_{\rm em}({\bm r})\cdot 
   {\bm \epsilon}_{+1} \, {\rm e} ^{{\rm i}{\mbi k} 
     \cdot {\mbi r} } |\tilde{\rm {N}}_{-{1\over 2}, M_T }\rangle
\label{Ahalf}
\end{equation}
where $k_W$ is the photon momentum at the photon point,
and the scalar helicity amplitude as
\begin{equation}
S_{1/2}={\zeta}\,\sqrt{2 \pi \alpha \over k_W}  \int \d {\bm r}  \, 
    \langle\tilde{\rm {R}}_{+{1\over 2}, M_T}|{J}^0_{\rm em} ({\bm r}) 
      \,\, 
     {\rm e} ^{{\rm i}{\mbi k} 
     \cdot {\mbi r} } |\tilde{\rm {N}}_{+{1\over 2}, M_T}\rangle\, .
\label{Shalf}
\end{equation}
Here ${J}^\mu_{\rm em}$ is the EM current derived from the
Lagrangian density (\ref{Lagrangian}):
\begin{equation}
  {J}^\mu_{\rm em} = 
    \sum_{i=1}^3 {\overline{q}}_i \gamma^{\mu}(i) \left( {1 \over 6} 
  + {1 \over 2} \tau_0{(i)} \right) {q}_i +
   \left(\vec{{\pi}} \times \partial ^\mu \vec{{\pi}} \right)_0 \;.
\label{Jem}
\end{equation}
The amplitudes (\ref{Ahalf}) and (\ref{Shalf}) contain a phase factor,
$\zeta$, determined by the sign of the decay amplitude into the nucleon 
and the pion.

The new term in (\ref{Rext}) {\em does not contribute}
to the nucleon-Roper transition amplitudes.
Namely, for an arbitrary EM transition operator $\hat{O}$ involving
only quarks and pions we can write
\begin{equation}
  \langle N|\tilde{a}_1 \hat{O}|N\rangle =
  \langle N|[\tilde{a}_1, \hat{O}]|N\rangle  = 0\;,  
\end{equation} 
because of (\ref{an}) and since the operators $\tilde{a}_n$ commute with $\hat{O}$.

A possible way to identify such a state would be to search for those
excited states for which the amplitudes are strongly reduced compared 
to those calculated in a model with only quark degrees of freedom.
There have been several attempts to calculate these amplitudes
in various models, such as in chiral quark models~\cite{AFGM,tiator},
models with explicit gluon degrees of freedom~\cite{Li} and
relativistic versions of the constituent quark model~\cite{Capstick}.
Unfortunately, the present status of theoretical prediction is
rather unclear because of a strong sensitivity of transition
amplitudes on small variation of the Roper wave function.
To understand the nature of the Roper remains a big challenge
for theoreticians as well as for experimentalists.

\end{document}